\documentclass[iop]{emulateapj}
\usepackage{amssymb}
\usepackage{apjfonts}
\usepackage{graphicx}
\usepackage{multirow}

\usepackage{amsmath}
\usepackage{epstopdf}
\usepackage{natbib} 
\usepackage{threeparttable} 

\slugcomment{prepared for ApJ}
\begin{document}
\title{Search for gamma-ray emission from the Sun during solar minimum with the ARGO-YBJ experiment}
\author{B. ~Bartoli\altaffilmark{1,2},
 P. ~Bernardini\altaffilmark{3,4},
 X. J. ~Bi\altaffilmark{5,6},
 Z. ~Cao\altaffilmark{5,6},
 S. ~Catalanotti\altaffilmark{1,2},
 S. Z. ~Chen\altaffilmark{5},
 T. L. ~Chen\altaffilmark{7},
 S. W. ~Cui\altaffilmark{8},
 B. Z. ~Dai\altaffilmark{9},
 A. ~D'Amone\altaffilmark{3,4},
 Danzengluobu\altaffilmark{7},
 I. ~De Mitri\altaffilmark{23,24},
 B. ~D'Ettorre Piazzoli\altaffilmark{1},
 T. ~Di Girolamo\altaffilmark{1,2},
 G. ~Di Sciascio\altaffilmark{10},
 C. F. ~Feng\altaffilmark{11},
 Zhaoyang Feng\altaffilmark{5},
 Zhenyong Feng\altaffilmark{12},
 W. ~Gao\altaffilmark{11},
 Q. B. ~Gou\altaffilmark{5},
 Y. Q. ~Guo\altaffilmark{5},
 H. H. ~He\altaffilmark{5,6},
 Haibing Hu\altaffilmark{7},
 Hongbo Hu\altaffilmark{5},
 M. ~Iacovacci\altaffilmark{1,2},
 R. ~Iuppa\altaffilmark{13,14},
 H. Y. ~Jia\altaffilmark{12},
 Labaciren\altaffilmark{7},
 H. J. ~Li\altaffilmark{7},
 Z. ~Li\altaffilmark{5,*}, 
 C. ~Liu\altaffilmark{5},
 J. ~Liu\altaffilmark{5},
 M. Y. ~Liu\altaffilmark{7},
 H. ~Lu\altaffilmark{5},
 L. L. ~Ma\altaffilmark{5},
 X. H. ~Ma\altaffilmark{5},
 G. ~Mancarella\altaffilmark{3,4},
 S. M. ~Mari\altaffilmark{15,16},
 G. ~Marsella\altaffilmark{3,4},
 S. ~Mastroianni\altaffilmark{2},
 P. ~Montini\altaffilmark{17},
 C. C. ~Ning\altaffilmark{7},
 L. ~Perrone\altaffilmark{3,4},
 P. ~Pistilli\altaffilmark{15,16},
 P. ~Salvini\altaffilmark{18},
 R. ~Santonico\altaffilmark{10,19},
 P. R. ~Shen\altaffilmark{5},
 X. D. ~Sheng\altaffilmark{5},
 F. ~Shi\altaffilmark{5},
 A. ~Surdo\altaffilmark{4},
 Y. H. ~Tan\altaffilmark{5},
 P. ~Vallania\altaffilmark{20,21},
 S. ~Vernetto\altaffilmark{20,21},
 C. ~Vigorito\altaffilmark{21,22},
 H. ~Wang\altaffilmark{5},
 C. Y. ~Wu\altaffilmark{5},
 H. R. ~Wu\altaffilmark{5},
 L. ~Xue\altaffilmark{11},
 Q. Y. ~Yang\altaffilmark{9},
 X. C. ~Yang\altaffilmark{9},
 Z. G. ~Yao\altaffilmark{5},
 A. F. ~Yuan\altaffilmark{7},
 M. ~Zha\altaffilmark{5},
 H. M. ~Zhang\altaffilmark{5},
 L. ~Zhang\altaffilmark{9},
 X. Y. ~Zhang\altaffilmark{11},
 Y. ~Zhang\altaffilmark{5},
 J. ~Zhao\altaffilmark{5},
 Zhaxiciren\altaffilmark{7},
 Zhaxisangzhu\altaffilmark{7},
 X. X. ~Zhou\altaffilmark{12},
 F. R. ~Zhu\altaffilmark{12},
 Q. Q. ~Zhu\altaffilmark{5}\\ (The ARGO-YBJ Collaboration)}


\altaffiltext{*}{Corresponding author: Zhe Li, lizhe@ihep. ac. cn; Songzhan Chen, chensz@ihep. ac. cn; Huihai He, hhh@ihep. ac. cn;D'Ettorre Piazzoli,B., dettorre@na. infn. it}
 \affil{ \altaffilmark{1}Dipartimento di Fisica dell'Universit\`a di Napoli
     ``Federico II'', Complesso Universitario di Monte
     Sant'Angelo, via Cinthia, 80126 Napoli, Italy. }
 \affil{\altaffilmark{2}Istituto Nazionale di Fisica Nucleare, Sezione di
     Napoli, Complesso Universitario di Monte
     Sant'Angelo, via Cinthia, 80126 Napoli, Italy. }
 \affil{\altaffilmark{3}Dipartimento Matematica e Fisica "Ennio De Giorgi",
     Universit\`a del Salento,
     via per Arnesano, 73100 Lecce, Italy. }
 \affil{ \altaffilmark{4}Istituto Nazionale di Fisica Nucleare, Sezione di
     Lecce, via per Arnesano, 73100 Lecce, Italy. }
 \affil{ \altaffilmark{5}Key Laboratory of Particle Astrophysics, Institute
     of High Energy Physics, Chinese Academy of Sciences,
     P. O. Box 918, 100049 Beijing, P. R. China. }
    \affil{ \altaffilmark{6}School of Physical Sciences, University of Chinese Academy of Sciences, 100049 Beijing, People's Republic of China. }
 \affil{ \altaffilmark{7}Tibet University, 850000 Lhasa, Xizang, P. R. China. }
 \affil{\altaffilmark{8}Hebei Normal University, 050024, Shijiazhuang
     Hebei, P. R. China. }
 \affil{ \altaffilmark{9}Yunnan University, 2 North Cuihu Rd., 650091 Kunming,
     Yunnan, P. R. China. }
 \affil{ \altaffilmark{10}Istituto Nazionale di Fisica Nucleare, Sezione di
     Roma Tor Vergata, via della Ricerca Scientifica 1,
     00133 Roma, Italy. }
 \affil{\altaffilmark{11}Shandong University, 250100 Jinan, Shandong, P. R. China. }
 \affil{\altaffilmark{12}Southwest Jiaotong University, 610031 Chengdu,
     Sichuan, P. R. China. }
   \affil{\altaffilmark{13}Dipartimento di Fisica dell'Universit\`a di Trento, via Sommarive 14, I-38123 Povo, Italy.}
      \affil{\altaffilmark{14}Trento Institute for Fundamental Physics and Applications, via Sommarive 14, I-38123 Povo, Italy.}
 \affil{ \altaffilmark{15}Dipartimento di Fisica dell'Universit\`a ``Roma Tre'',
     via della Vasca Navale 84, 00146 Roma, Italy. }
 \affil{ \altaffilmark{16}Istituto Nazionale di Fisica Nucleare, Sezione di
     Roma Tre, via della Vasca Navale 84, 00146 Roma, Italy. }
   \affil{\altaffilmark{17}Dipartimento di Fisica dell'Universit\`a di Roma "La Sapienza" and INFN Sezione di Roma, piazzale Aldo Moro 2, I-00185 Roma, Italy. }
 \affil{ \altaffilmark{18}Istituto Nazionale di Fisica Nucleare, Sezione di Pavia,
     via Bassi 6, 27100 Pavia, Italy. }
 \affil{ \altaffilmark{19}Dipartimento di Fisica dell'Universit\`a di Roma
     ``Tor Vergata'', via della Ricerca Scientifica 1,
     00133 Roma, Italy. }
 \affil{\altaffilmark{20}Osservatorio Astrofisico di Torino dell'Istituto Nazionale
     di Astrofisica, via P. Giuria 1, 10125 Torino, Italy. }
 \affil{ \altaffilmark{21}Istituto Nazionale di Fisica Nucleare,
     Sezione di Torino, via P. Giuria 1, 10125 Torino, Italy. }
 \affil{ \altaffilmark{22}Dipartimento di Fisica dell'Universit\`a di
     Torino, via P. Giuria 1, 10125 Torino, Italy. }
 \affil{ \altaffilmark{23}Gran Sasso Science Institute(GSSI), Via Iacobucci 2, 67100 L'Aquila, Italy.}
 \affil{ \altaffilmark{24}Istituto Nazionale di Fisica Nucleare (INFN), Laboratori Nazionali del Gran Sasso, 67100, Assergi, L'Aquila, Italy.}

 \begin{abstract}
The hadronic interaction of cosmic rays with solar atmosphere can produce high energy gamma rays. The gamma-ray luminosity is correlated both with the flux of primary cosmic rays and the intensity of the solar magnetic field. The gamma rays below 200 GeV have been observed by $Fermi$ without any evident energy cutoff. The bright gamma-ray flux above 100 GeV has been detected only during solar minimum. The only available data in TeV range come from the HAWC observations, however outside the solar minimum. The ARGO-YBJ dataset has been used to search for sub-TeV/TeV gamma rays from the Sun during the solar minimum from 2008 to 2010, the same time period covered by the Fermi data. A suitable model containing the Sun shadow, solar disk emission and inverse-Compton emission has been developed, and the chi-square minimization method was used to quantitatively estimate the disk gamma-ray signal. The result shows that no significant gamma-ray signal is detected and upper limits to the gamma-ray flux at 0.3$-$7 TeV are set at 95\% confidence level. In the low energy range these limits are consistent with the extrapolation of the Fermi-LAT measurements taken during solar minimum and are compatible with a softening of the gamma-ray spectrum below 1 TeV. They provide also an experimental upper bound to any solar disk emission at TeV energies. Models of dark matter annihilation via long-lived mediators predicting gamma-ray fluxes > $10^{-7}$ GeV $cm^{-2}$ $s^{-1}$ below 1 TeV are ruled out by the ARGO-YBJ limits.
\end{abstract}


\section{Introduction}
There are several mechanisms for high energy gamma-ray emission from the solar region. The Sun can emit electromagnetic radiation extending from radio to gamma rays during solar flare, which is likely associated with the interaction of flare-accelerated particles in the solar atmosphere. Up to now, dozens of solar flares have been detected by The $Fermi$ Large Area Telescope (LAT) with gamma-ray emission above 100 MeV \citep{acke14}. The maximum energy observed up to now is about 4 GeV \citep{ajel14}. Another plausible mechanism is the self-annihilation of dark matter, i.e. heavy Weak Interacting Massive Particles (WIMPs), which may accumulate near the Sun when they lose energy upon scattering and are gravitationally captured. The Sun has recently been proposed as an intense source of high energy gamma rays from dark matter annihilation via long-lived mediators \citep{arina17,leane17}. Fluxes comparable or greater than the Crab Nebula flux are predicted in some of the proposed models. Apart from this conjecture, the most important astrophysical mechanism for steady solar gamma-ray production is the interaction of cosmic rays with solar matter and photons, that has been definitely detected by $Fermi$-LAT with maximum energy up to 200 GeV \citep{abdo11,kenny16,tang18}.

The gamma-ray emission from the solar disk due to CR cascades in the solar atmosphere is denoted as disk component. This secondary gamma-ray produced by the hadronic interaction of cosmic ray with the solar surface was firstly proposed by \cite{dolan65}. While, only upper limits were obtained by early measurements over the range 20 keV to 10 MeV \citep{peter66}. A detailed theoretical model for gamma rays from the collision of cosmic ray with the solar atmosphere is presented by \cite{seckel91}. The predicted gamma-ray flux at energies from 10 MeV to 10 GeV has a large uncertainty, being sensitive to the assumptions about the cosmic ray transport in the magnetic field near the Sun. Gamma rays from the Sun were firstly detected by the Energetic Gamma-ray Experiment Telescope (EGRET) \citep{orlando08}. The measured flux from 100 MeV to 2 GeV was within the range of the theoretical predictions. The $Fermi$ collaboration \citep{abdo11} reported the detection of high-energy gamma rays at 0.1-10 GeV from the quiescent Sun using the first 1.5 years data. However, the measured solar disk emission flux was about a factor of seven higher than that predicted about this disk component by a "nominal" model \citep{seckel91}. This mismatch motivated \cite{kenny16} to analyze 6 years of public $Fermi$-LAT data. The obtained gamma-ray spectrum follows a simple power-law shape ($\alpha$=-2.3) in 1$-$100 GeV without any evident high energy cutoff. For the flux in 1$-$10 GeV, a significant time variation of the solar-disk gamma-ray flux which anticorrelates with solar activity was discovered, suggesting that the solar magnetic field would play an important role. An updated analysis with 9 years of $Fermi$-LAT data, from 2008-8-7 to 2017-7-27, \cite{tang18} confirms these results and extends the gamma-ray spectrum up to > 200 GeV. Notably, the bright gamma-ray flux above 100 GeV is dominant only during solar minimum at the end of Cycle 23 \citep{linden18}. The HAWC measurements in periods of high solar activity may support these findings \citep{albert18}. Data collected from November 2014 to December 2017, the second half of solar cycle 24, have been used to set strong upper limits on the flux of 1-100 TeV gamma rays from the solar disk, about 10\% of the maximum gamma-ray flux estimated by \cite{linden18}. The HAWC 95\% upper limit at 1 TeV is about 13\% of the flux extrapolated from the solar minimum Fermi-LAT gamma-ray spectrum.

Besides the disk component produced by the hadronic interaction of cosmic rays with the solar atmosphere, the interaction of cosmic ray electrons and positrons with solar photons can also produce high energy gamma rays via the inverse Compton (IC) scattering. The IC component and the corresponding flux were predicted by \cite{moskal06} and \cite{orlando08}. Lately, the IC component was clearly observed by $Fermi$-LAT \citep{abdo11} in 0.1$-$10 GeV and the measured flux is in
good agreement with the theoretical prediction. The IC component appears as an extended halo centered on the Sun direction with extension radius up to about 20$^{\circ}$. Its spectrum follows a power-law shape up to about 10 GeV.

According to \cite{seckel91}, the disk flux is enhanced by magnetic effects that gradually
reduce above a critical proton energy between 3$\times$10$^2$ GeV and 2$\times$10$^4$ GeV. Thus,gamma rays above the corresponding critical energy around 1 TeV
should be unaffected by the solar magnetic field. Recently \citep{zhou17} have estimated the solar disk gamma-ray flux up to 1000 TeV
without taking into account the effect of the solar magnetic field. The
predicted flux is at least one order of magnitude lower than that measured
by $Fermi$-LAT. Hence, the observed gamma-ray flux should have been
significantly boosted by the solar magnetic field and the spectrum would
change around the critical energy. A dependence on the phase of the solar
cycle can be also anticipated. The \citep{kenny16} analysis of $Fermi$-LAT
data shows that the flux of low energy solar disk gamma rays
anticorrelates with solar activity. However, solar activity could affect
even the flux of high energy gamma rays. Indeed, in the Tibet AS$\gamma$
observations of the Sun shadow \citep{ameno13}, the cosmic rays at
10 TeV are little affected during the solar quiet phase, while they are
largely blown away from the Sun during the solar active phase.
The ARGO-YBJ collaboration reported on the rigidity dependent
variation of the Sun shadow in the rigidity range 0.4-200 TV using data
collected in the years 2008-2012 when the solar activity varied from the
minimum to the maximum \citep{chen17}. The number of deficit events
and the shape of the Sun shadow turn out to be strictly correlated with
the solar activity. The characteristic rigidity, corresponding to a
deficit ratio ( observed deficit/expected deficit in the absence of
magnetic field) of 50\%, ranges from 1 TV to 16 TV during this period.
Likewise, the critical energy should vary across 1 TeV during a solar
cycle. New observations above 100 GeV will provide important clues for the gamma-ray production from the Sun, the magnetic field intensity and the corresponding cosmic ray modulation near the Sun. The large gamma-ray fluxes predicted by some models of dark matter annihilation outside the Sun could be efficiently probed by observations in the sub-TeV/TeV range.

To extend solar gamma-ray observations into the VHE range, i.e.>100 GeV, large ground-based detectors are needed. There are two main classes of ground-based gamma-ray detectors: the Extensive Air Shower (EAS) arrays and the Imaging Atmospheric Cherenkov Telescopes (IACTs). IACTs have an excellent sensitive for gamma rays, while they can't work in day time making impossible to observe gamma rays from the bright Sun. Therefore, the EAS array is the only choice for the Sun observation. The ARGO-YBJ detector is an EAS array with a large field of view (FOV) and can observe gamma rays at an energy threshold of $\sim$300 GeV. Previously, the ARGO-YBJ collaboration achieved some important results in TeV gamma-ray observations, e. g. flaring activity of AGNs (\citealp{barto12a,barto16}), northern sky survey \citep{barto13}, Cygnus Cocoon\citep{barto14}, Galactic plane diffuse gamma-ray emission \citep{barto15}.
The ARGO-YBJ sensitivity for a gamma point
source is about 24\% Crab flux, however the large mismatch between the
observed flux at low energies and the theoretical predictions, and the chance
to probe new physics, motivate an
observational search for sub-TeV/TeV gamma rays from the Sun direction .
For this purpose we have carried out this search using 3 years of
ARGO-YBJ data collected from 2008 to 2010.Only the solar disk component is concerned in this work.

The paper is structured as follows. The ARGO-YBJ experiment and the
selection of the data set used in the analysis are presented in Section 2
. The analysis procedure is described in Section 3. The results
concerning the ARGO-YBJ sensitivity, the upper limits on the gamma-ray
flux from the solar disk and the systematic uncertainties are detailed in
Section 4. These results and their implications are discussed in Section
5 and summarized in Section 6.

\section{Experiment and Data Selection}
\subsection{The ARGO-YBJ experiment}
The ARGO-YBJ experiment, installed in the Cosmic Ray Observatory of Yangbajing (Tibet, P.R.China) at an altitude of 4300 m above sea level, consists of a single layer of Resistive Plate Chambers (RPCs) with a 74$\times$78 m$^2$ carpet (93\% coverage) in the center surrounded by a partially instrumented ring (20\% coverage) extending the whole area to 100$\times$110 m$^2$. More details about the detector can be found in \cite{aiel06}.
The trigger rate is 3.5 kHz with a dead time of 4\% and an average duty-cycle higher than 86\%.
It can detect air shower induced by gamma rays and cosmic-rays from about 300 GeV to PeV energies. ARGO-YBJ, with a FOV of 2 sr, can monitor the whole overhead sky up to zenith angle >50$^{\circ}$ \citep{dettorre13}. This property makes possible to observe gamma rays from the Sun during most of the days in one year.

\subsection{Energy Bin and Angular Resolution}
The energy and direction reconstruction are based on Monte Carlo simulation samples. A full Monte Carlo simulation process is implemented
using CORSIKA 6.502 \citep{heck98} for extensive air showers and GEANT4-based code G4argo \citep{guo10} for detector response. The energy range of the simulated showers is from 10 GeV to 1 PeV, all of these simulated events are used for the energy reconstruction and to estimate the angular resolution, to evaluate the exposure and obtain the flux. The primary true energy (E$_{true}$) of a shower is mainly related to the number of secondary particles recorded by the detector, and also to the incident zenith angle and core position, and this work has been discussed in detail in \cite{barto18}. Fig.1 shows the relation between the reconstructed energy (E$_{rec}$) and the primary true energy (E$_{true}$) for gamma rays. In Fig.1 the error bar on E$_{rec}$ is the width of the energy bin while the error bar on E$_{true}$ is the RMS of the distribution. The events with reconstructed energy from 0.32 TeV to 10 TeV are divided into 6 energy bins with intervals of 0.25 in log space, centered at 0.28 TeV, 0.56 TeV, 1.01 TeV, 2.02 TeV,3.76 TeV and 6.53 TeV. For each energy bin, the angular resolution ($\sigma_{res}$) of ARGO-YBJ for cosmic rays and gamma rays are shown in Table 1. The quoted angular resolution is the standard deviation of the two-dimensional Gaussian distribution which fits the detector point spread function (PSF).

\subsection{Sun Observation and Data Selection}
The ARGO-YBJ data taking in its full configuration covers the period 2007 November-2013 February. This period encloses different epochs, the years 2008-2009 of very low solar activity ( the declining phase of Cycle 23), the year 2010 corresponding to the ascending phase of Cycle 24, and the period 2011-2012 of increasing solar activity preceding the Northern polar field sign switch in 2012 November \citep{sun15}. This last period is characterized by complexity and variability of the solar and interplanetary magnetic fields. $Fermi$-LAT
data shows a strong dependence of the gamma-ray flux on the solar activity, and a spectrum extending up to 200 GeV only during solar minimum. \cite{zhu15} analyzed the whole ARGO-YBJ data sample, showing that in this period of intense solar activity the Sun shadow of 5 TeV cosmic rays, whose imprint features the excess map built from data to search for gamma rays from the solar disk, has an erratic shape since the particle deflections are highly randomized. The effect is stronger for sub-TeV cosmic rays. Modeling this deficit is not a straightforward matter. Consequently, only data collected in the 2008-2010 years have been analyzed in the present work.

Since no specific selection is applied to data, the events from the Sun
direction are dominated by the cosmic ray background. Thus, the gamma-ray
signal can be detected as an excess above this background. However, the
current case is not alike to the search of high energy gamma rays from
point sources. Indeed the Sun blocks galactic cosmic rays casting a shadow
that appears as a deficit region in the event sky map. The Sun shadow has
been widely detected by Tibet AS$\gamma$ \citep{ameno13} and ARGO-YBJ \citep{zhu11,chen17} experiments. The shape and
intensity of the shadow are affected by all the magnetic fields between
the Sun and the Earth, that is the solar magnetic
field, the interplanetary magnetic field and the geomagnetic field. As a
consequence, the number of deficit events is correlated with the variable
solar magnetic field. These deficit events would counteract any excess
signal from the Sun direction. However, due to the bending effect of the
magnetic fields, the Sun shadow is displaced with respect to the Sun
position. The offset is rigidity dependent, decreasing with increasing
rigidity. Therefore, at low rigidities, the center of the Sun shadow and
the solar gamma-ray signal are not completely overlapped in the sky map.
To exploit this misalignement, only shower with energy less than 10 TeV
have been used in the analysis.

To estimate the performance of the ARGO-YBJ detector for the observation of air showers from the Sun direction,the Sun is traced continuously over one full observation year, i.e. 365 transits. Fig.2 shows the ideal daily allowable Sun observation time during the full ARGO-YBJ operation time from 2008 January to 2010 December. The actual observation time was 4393. 2 hours.To avoid the influence of other sources, the periods when the Sun is close to the Moon and to the known gamma-ray sources listed in \citep{barto13} with space angle less than 8$^{\circ}$ are excluded. With this selection, the observation time on Sun is reduced to 3891.7 hours.

For the analysis presented in this paper, events are selected according to the following cuts: (1) zenith angle $\theta$<50$^{\circ}$; (2)
the distance between the shower core position and the carpet center less than 100 m; (3) the time spread of the shower front in the
conical fit defined in Equation (1) of \cite{aiel09} less than 100 ns$^2$. These cuts, only slightly different from the ones used for studies in gamma-ray astronomy, \cite{barto13a} and \cite{barto15}, optimize the observation of gamma-rays from the Sun.

\begin{figure}[h!]
\centering
\includegraphics[width=3.5in,height=3.0in]{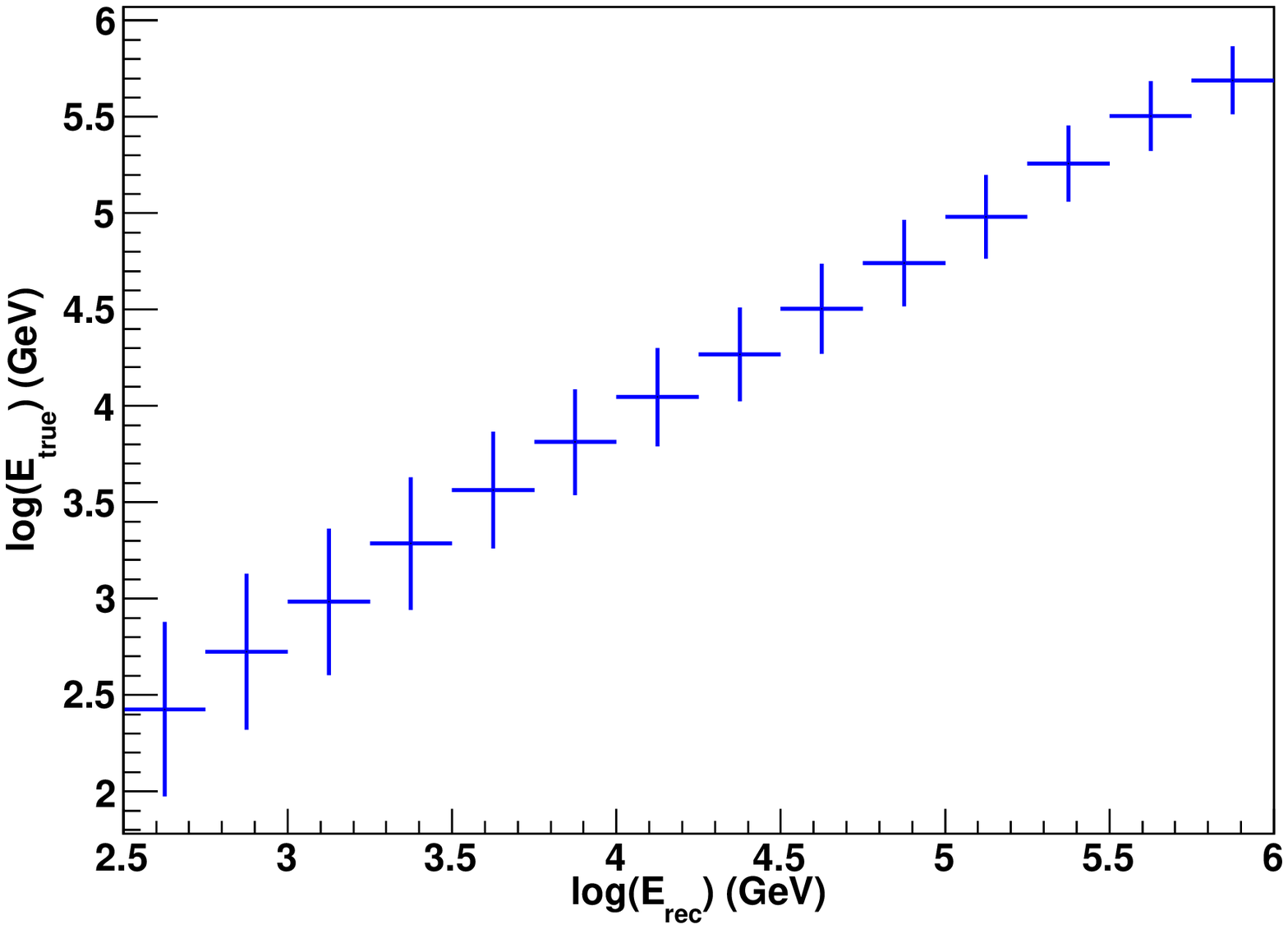}
\begin{center}
\hspace{5mm}{\textbf{Figure 1.}The correlation between the reconstructed energy and the primary energy for gamma rays. The error bar on E$_{rec}$ is the width of the energy bin while the error bar on E$_{true}$ is the RMS of the distribution.}
\end{center}
\label{fig2}
\end{figure}

\begin{figure*}
\centering
\includegraphics[width=7in,height=2.4in]{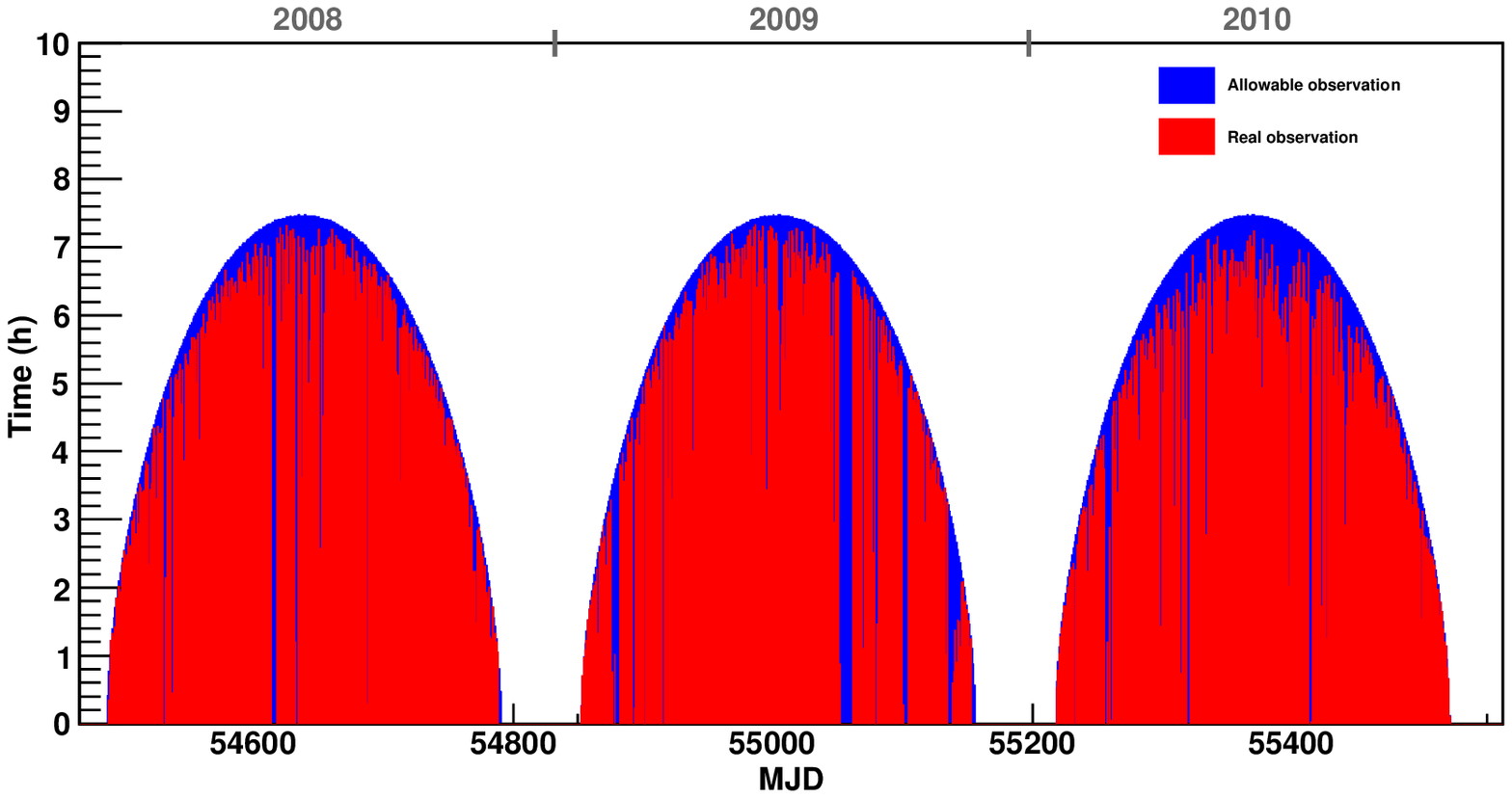}
\begin{center}
\hspace{5mm}{\textbf{Figure 2.} Daily allowable Sun observation time for ARGO-YBJ during the period from 2008 January to 2010 December. The zenith angle of the Sun is less than 50$^{\circ}$.}
\end{center}
\label{fig1}
\vspace*{0.5cm}
\end{figure*}

\begin{table}[ht]
\caption{The median energy of E$_{true}$ for gamma rays and the angular resolution for cosmic rays and gamma rays} 
\centering 
\begin{tabular}{lccc}
\hline
log(E$_{rec}$) (GeV)& E$_{true}$ (TeV) &$\sigma_{res}$ &$\sigma_{res}$  \\
    &     &  for cosmic rays  & for gamma rays  \\
\hline
2.50$-$2.75 & 0.28 &2.52$^{\circ}$	&1.84$^{\circ}$\\
2.75$-$3.00	& 0.56 &1.77$^{\circ}$	&1.35$^{\circ}$\\
3.00$-$3.25	& 1.01 &1.52$^{\circ}$	&1.30$^{\circ}$\\
3.25$-$3.50	& 2.02 &1.18$^{\circ}$	&1.00$^{\circ}$\\
3.50$-$3.75	& 3.76 &0.83$^{\circ}$	&0.74$^{\circ}$\\
3.75$-$4.00 & 6.53 &0.63$^{\circ}$	&0.60$^{\circ}$\\
\hline
\end{tabular}
\label{table1} 
\end{table}

\section{Analysis Methods}
\subsection{Significance Analysis}
The 40$^{\circ}$$\times$40$^{\circ}$ sky region in celestial coordinates (right ascension R.A., declination DEC) centered on the Sun is
divided into a grid of 0.1$^{\circ}$$\times$0.1$^{\circ}$ pixels and filled with the detected
events. The number of events in each pixel is denoted by n. In order to obtain the excess of gamma-induced showers in each bin, the "direct
integral method" \citep{fleys04} is adopted to estimate the number
of background events in each pixel (denoted by $b$).
It is worth to note that the regions with a radius of 8$^{\circ}$ centered on the Sun, Moon and the known gamma-ray sources listed in \citep{barto13} are excluded from the background estimation. To cope with a faint tail of the Sun shadow and Moon shadow observed at low energies, the excluded region around the Sun and Moon has been shifted by 3$^{\circ}$ in the right ascension East-West direction. Both the event and the
background maps are smoothed according to the angular resolution \citep{barto13}. Then the background map has been subtracted to the event map
obtaining the source map. The Li-Ma method \citep{Li83} is used to
estimate the significance of excess or deficit in each pixel, providing the
significance map.

The significance maps for the six energy bins are shown in Fig.3. The
Sun shadow is well evident in each map, with an intensity increasing with
the energy. As expected, the shadow profile is broadened and the shadow
center more shifted as the median energy decreases. No significant excess signal is detected at the Sun position.

\begin{figure*}
\centering
\includegraphics[width=6in,height=4in]{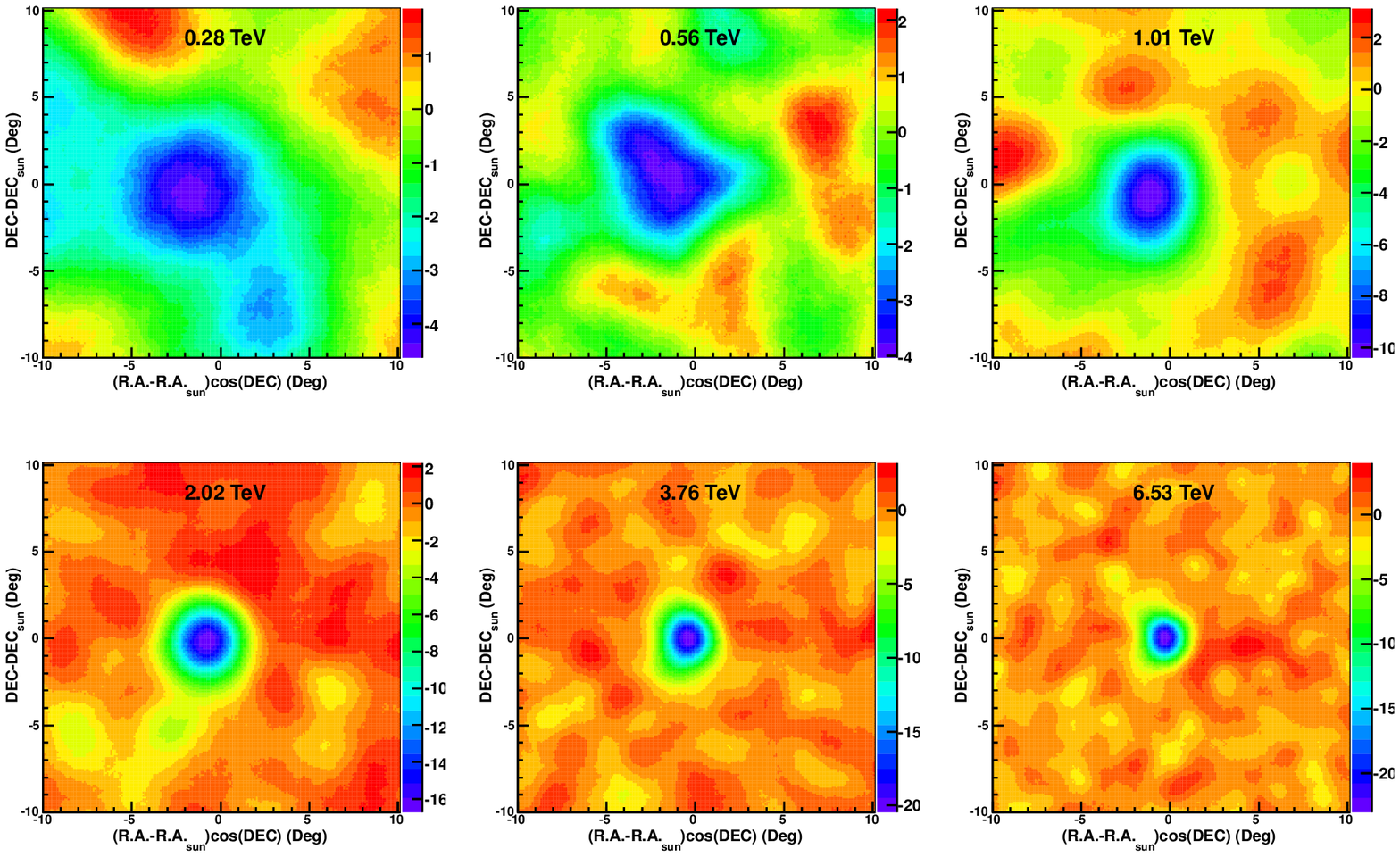}
\begin{center}
\hspace{5mm}{\textbf{Figure 3.} Significance maps for six energy bins. The coordinates are R.A. and DEC centered on the Sun position($R.A._{sun}$,$DEC_{sun}$). The color scales are different for each energy bin.}
\end{center}

\label{fig3}
\vspace*{0.5cm}
\end{figure*}

\subsection{Source Model}
As discussed in Sect.2, the presence of the shadow may offset the gamma-ray signal also at low energies where the shadow does not completely overlap the solar disk. To extract the signal from the Sun direction, a function was developed to model the source map content
\begin{equation}
 F(i)=s_i^+ + s^+_i,_{IC} -s_i^-
\end{equation}

where $s_i^+$ and $s^{+}_i,_{IC}$ are the expected number of events in the $i_{th}$ pixel from solar disk photons and IC photons, respectively, and $s_i^-$ is the expected cosmic ray deficit number in each pixel. The solar disk emission is centered on the Sun and modeled as the PSF for gamma rays (Table 1). The observed Sun shadow is approximately described by a two dimensional Gaussian distribution with
\begin{equation}
\sigma= \sqrt{\sigma_{det}^2+ \sigma_{sun}^2}
\end{equation}
where $\sigma_{det}$ is the detector resolution for cosmic rays (Table 1) and $\sigma_{sun}$ measures the intrinsic angular spread of the shadow. The model contains 5 free parameters, they are the total number of events from the solar disk, the position (right ascension and declination) of the peak of the deficit, the cosmic ray deficit number in the region of interest (ROI) and the parameter $\sigma_{sun}$. The IC component does not contain free parameters, we adopt the flux calculated by \cite{zhou17} with the intensity falling linearly with the angular distance from the center of the Sun \citep{abdo11}. Also this component is spread out according to the PSF for gamma rays. Therefore, a small correction is applied to taken into account a tiny contribution to the background from the IC halo. Thus, for each energy bin we have minimized the chi-square function
\begin{equation}
\chi^{2}=\sum_{i}^{ROI}[\frac{(s_i^+ +s^+_i,_{IC} - s_i^-) - (n_i - b_i )}{\sqrt{n_i + \alpha b_i}}]^2
\end{equation}

where $\alpha$ is the on to off-source time ratio and $\sqrt{n_i + \alpha b_i}$ is the uncertainty of ($n_i$-$b_i$)\citep{Li83}. The MINUIT/MINOS package \citep{james75} has been used to minimize the chi-square function of eq. (3) with respect to all model parameters.

Since the angular resolution for gamma rays is energy dependent and the peak position of the deficit is gradually shifted with decreasing energy while the deficit distribution spreads over
a wider angular range, the region of interest (ROI) is inevitably different for each energy bin. The adopted dimensions guarantee the full containment of gamma rays from the solar disk,
whose angular distribution is dictated by the detector angular resolution, and contains the main portion of the Sun shadow, taking into account the irregular shape of the shadow
and excluding part of the long tail observed at low energies. All ROIs are inside the excluded region around the Sun not involved in the background estimation.
 The dimensions of the ROI for each energy are reported in Table 2.

\begin{table*}[ht]
\centering
\caption{The ROI, the position of the Sun shadow peak, the upper limits on the flux and on the energy flux, and the systematic uncertainty affecting the upper limits are reported for each energy bin}
\centering 

\begin{tabular}{ccccccccc}
\hline
{E$_{true}$} \multirow{2}{*} &\multicolumn{2}{c}{ROI} & %
  \multicolumn{2}{c}{deficit peak position} &FluxUpperLimit &$\chi$$^{2}$/n.d.f  &E$^{2}$ $\times$ FluxUpperLimit  &systematic uncertainty \\
\cline{1-9}
 \cline{1-9}
TeV & DEC(deg) & R.A(deg) & DEC(deg) & R.A.(deg) & (GeV$^{-1}$cm$^{-2}$s$^{-1}$)  &   & (GeV cm$^{-2}$s$^{-1}$) & \\
\hline
0.28 &(-6,6) &(-9,6) &0.03$\pm$0.73  &-1.80$\pm$0.67	  &3.50e-13 &0.972  &2.75e-8  &44\% \\
0.56 &(-5,7) &(-7,5) &0.56$\pm$0.41  &-1.82$\pm$0.41	   &7.66e-14 &0.997  &2.40e-8  &24\% \\
1.01 &(-6,5) &(-5,5) &-0.67$\pm$0.15  &-1.24$\pm$0.13 	 &1.62e-14 &0.994  &1.65e-8  &30\% \\
2.02 &(-5,4) &(-5,4) &-0.14$\pm$0.07  &-0.74$\pm$0.15 	  &1.47e-14 &0.986  &5.99e-8  &6\% \\
3.76 &(-4,4) &(-4,4) &-0.02$\pm$0.04  &-0.46$\pm$0.03 	  &1.67e-15 &1.031  &2.36e-8  &5\% \\
6.53 &(-3,3) &(-3,2) &0.01$\pm$0.02  &-0.21$\pm$0.06 	  &1.37e-15 &1.016  &5.86e-8  &5\% \\

\hline
\end{tabular}
\begin{tablenotes}
\footnotesize
\item[1]NOTES:the equatorial coordinates R.A. and DEC are centered on the Sun position; the flux upper limits are at 95\% C.L.
\end{tablenotes}

\label{table2} 
\vspace*{0.5cm}
\end{table*}

\section{Results}
The observed event counts ($n_{i}$-$b_{i}$) projected on the right ascension axis centered on the Sun are shown in Fig.4 for each energy bin. The data are summed in the declination bands reported in Table 2. The contribution of each component of the source model is also shown. A slight displacement in declination of the deficit peak (Table 2), anyway compatible with the statistical uncertainty, can be attributed to the effect of the interplanetary magnetic field \citep{ameno00}.
 At low energies the Sun shadow is weak and shifted with respect to the Sun position, allowing a better sensitivity as compared to that reached at energies greater
 than 1 TeV where the shadow largely overlap the solar disk.

\begin{figure*}
\centering
\begin{minipage}[l]{0.3\textwidth}
 \includegraphics[height=4.7cm,width=6.2cm]{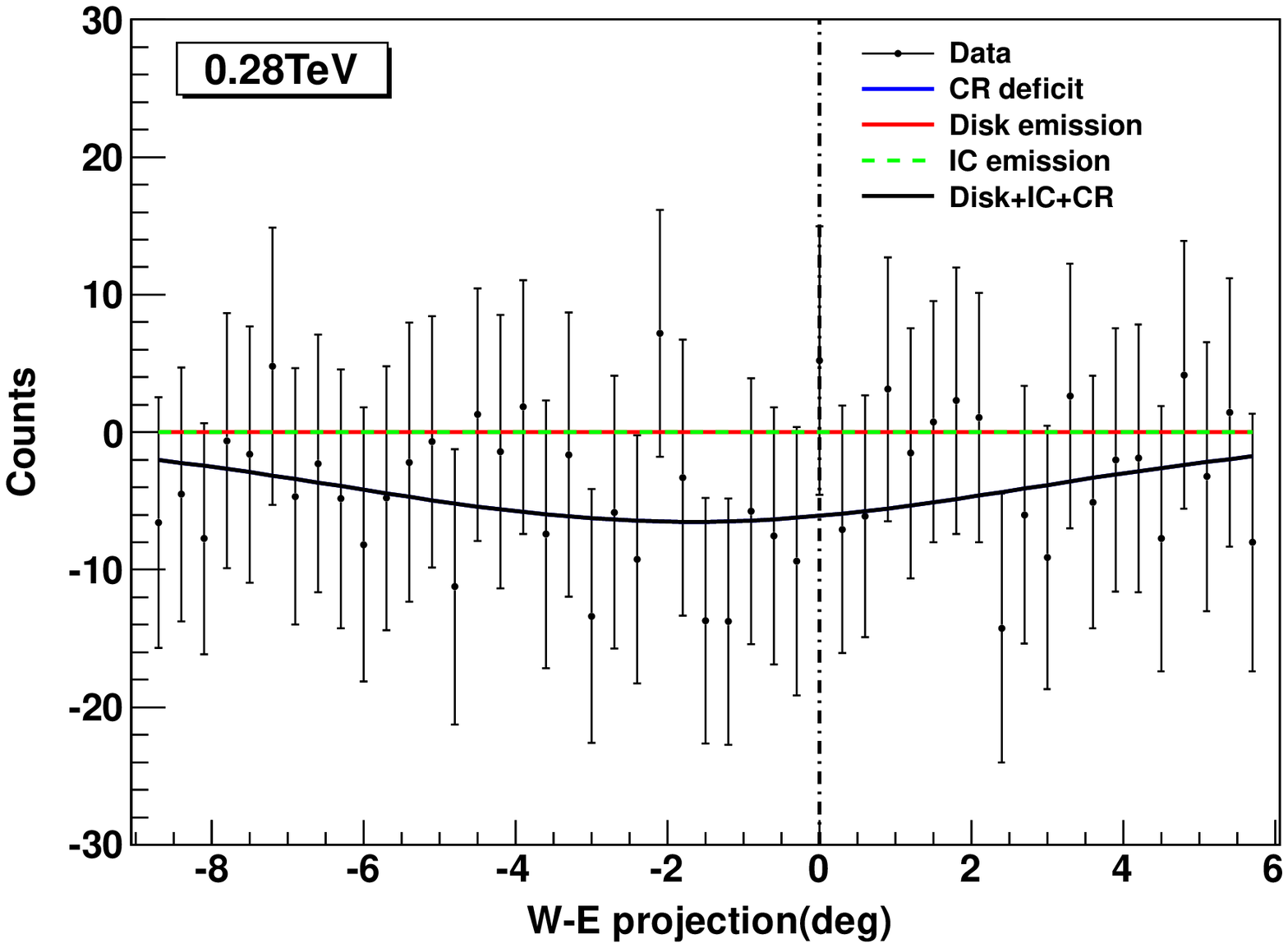}
 \end{minipage}%
 \hspace{5ex}
\begin{minipage}[l]{0.3\textwidth}
  \includegraphics[height=4.7cm,width=6.2cm]{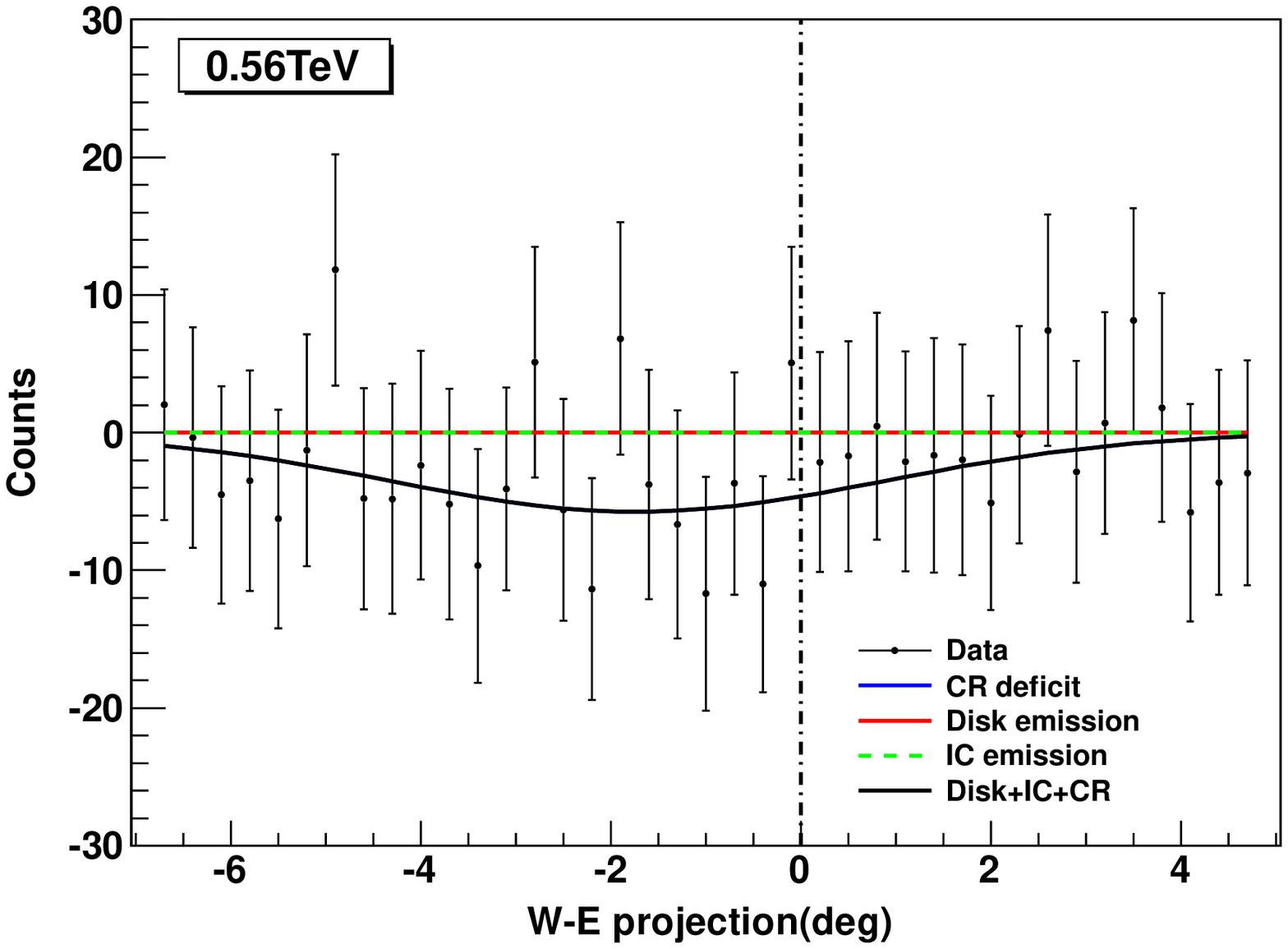}
\end{minipage}
\hspace{5ex}
\begin{minipage}[l]{0.3\textwidth}
\includegraphics[height=4.7cm,width=6.2cm]{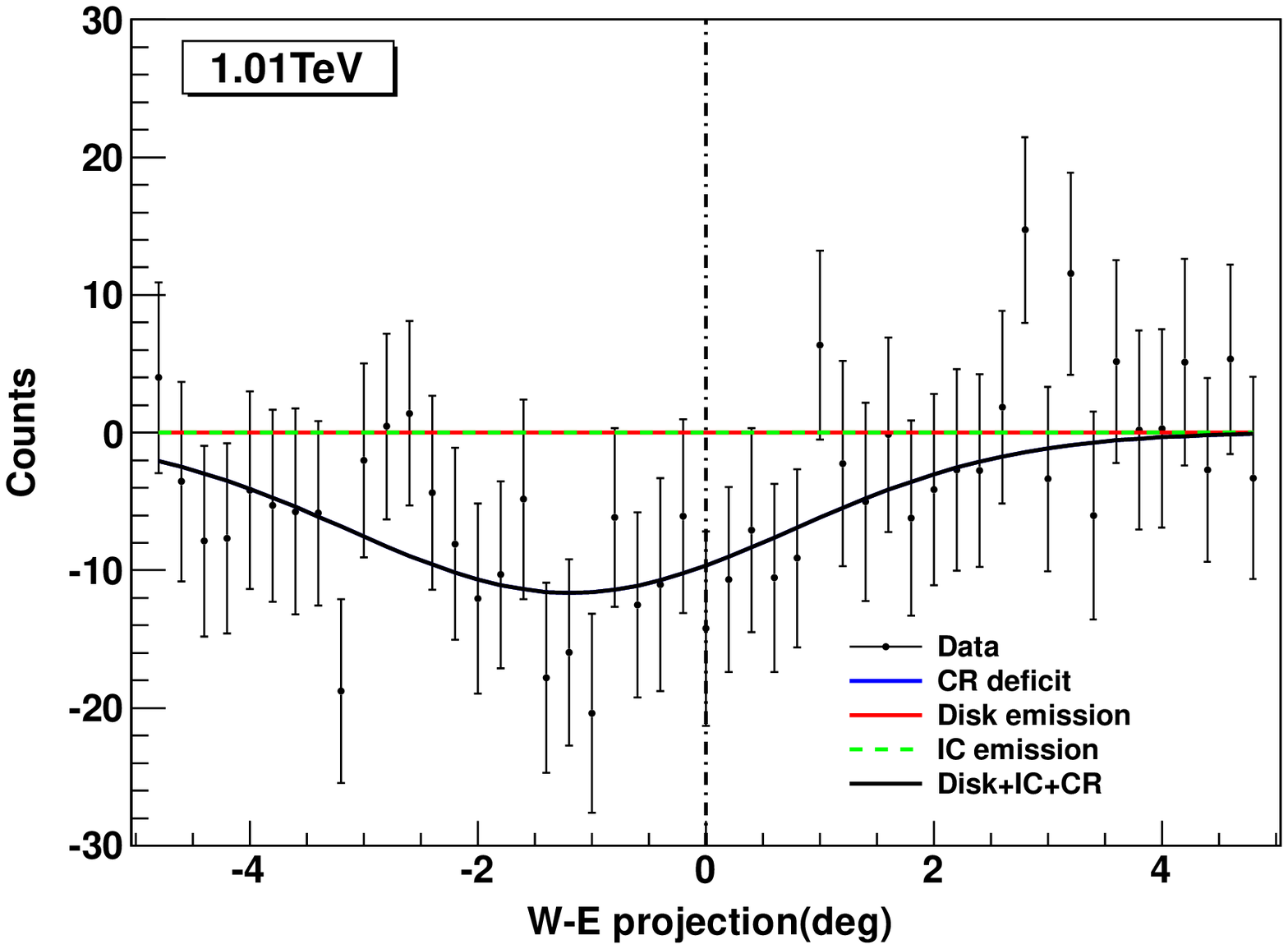}
\end{minipage}%
\begin{minipage}[l]{0.3\textwidth}
\includegraphics[height=4.7cm,width=6.2cm]{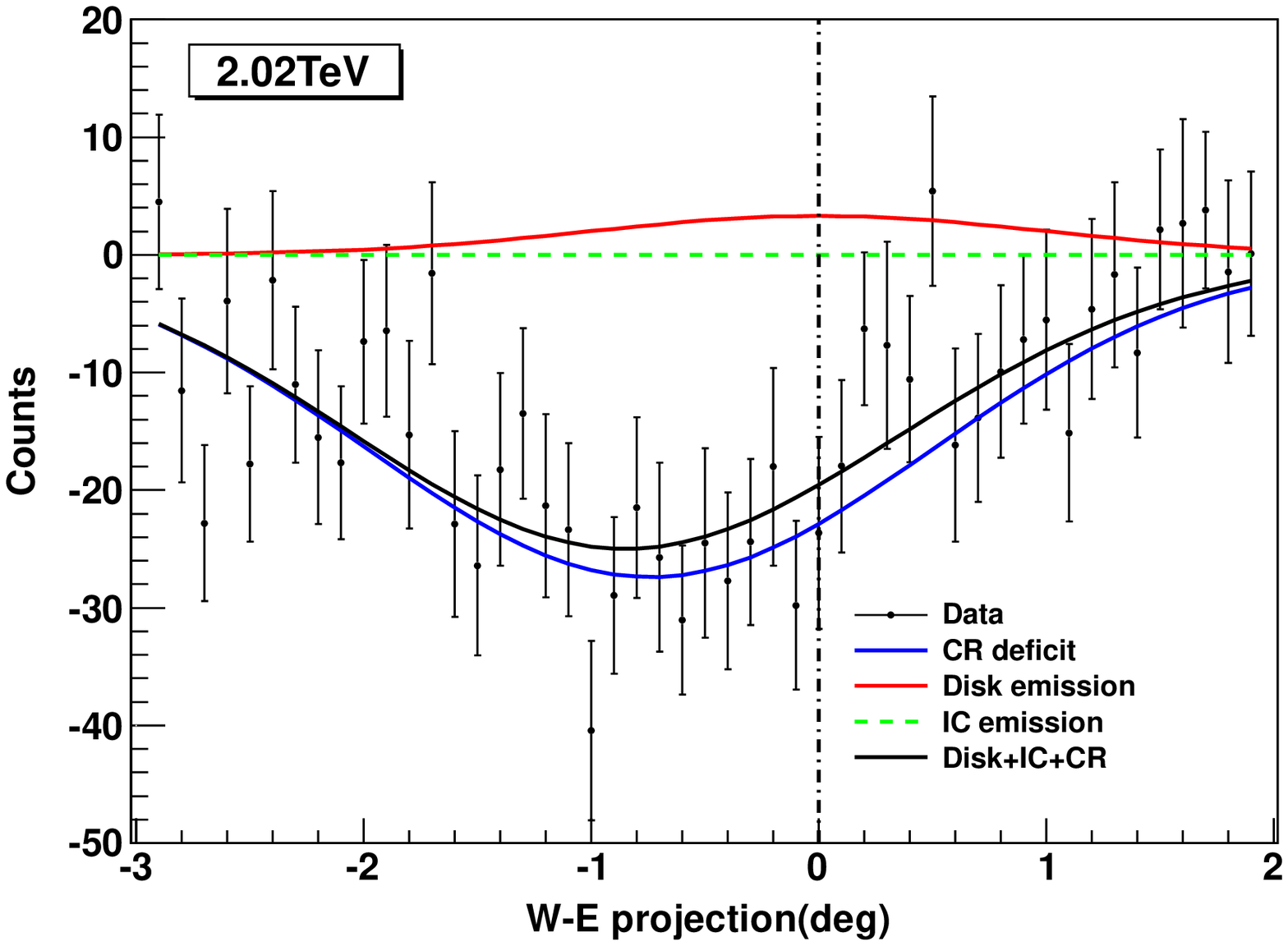}
\end{minipage}
\hspace{5ex}
\begin{minipage}[l]{0.3\textwidth}
\includegraphics[height=4.7cm,width=6.2cm]{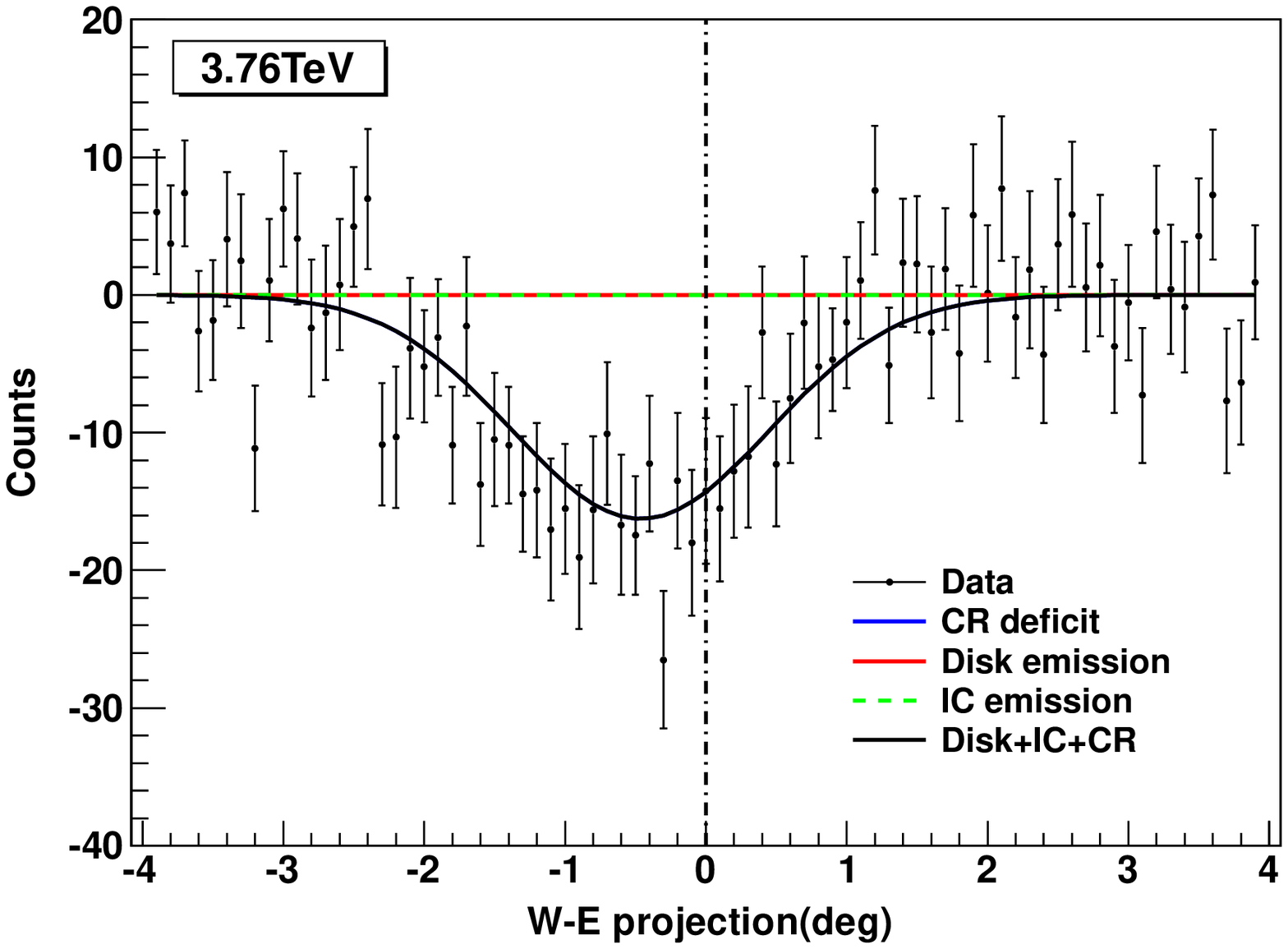}
\end{minipage}%
\hspace{5ex}
\begin{minipage}[l]{0.3\textwidth}
\includegraphics[height=4.7cm,width=6.2cm]{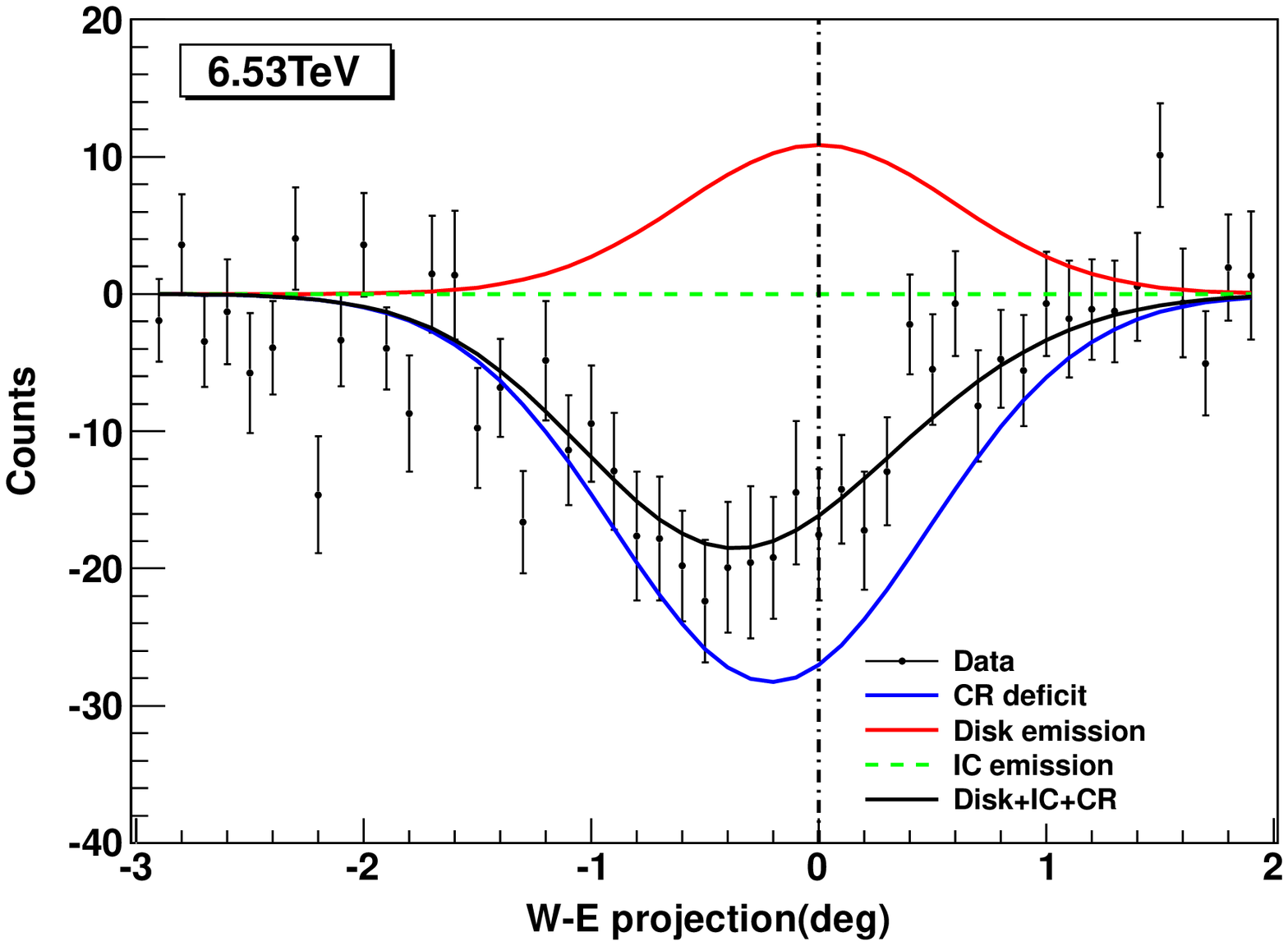}
\end{minipage}
\begin{center}
\hspace{5mm}{\textbf{Figure 4.}The ($n_{i}$-$b_{i}$) counts projected on the right ascension axis centered on the Sun for six energy bins. The data are summed in the declination band  reported in Table 2. Each component of the function $F(i)$ describing the source model is shown. The cosmic ray deficit (blue line) and the disk emission (red line) are the fit results. The IC component (green dashed line) has been taken from \cite{zhou17}.}
\end{center}
\label{fig3}       
\end{figure*}

\subsection{Upper Limit of Solar Gamma-ray}
Since no significant number of gamma-ray events is found from the solar disk direction in each energy bin, only a 95\% upper limit to the gamma-ray flux can be set by varying chi-square of 2.71 \citep{james75}. The upper limits at the median energy of each energy bin are given in Table 2.

The 95\% upper limit on the gamma-ray flux from the Sun at 0.28 TeV is about 37\% of the Crab Nebula flux
 measured by ARGO-YBJ \citep{barto15a}, and about 50\% at 0.56 TeV and 1.01 TeV. The limits found at higher energies are much less constraining
 being strongly affected by the Sun shadow. The last columns of Table 2 give the gamma-ray energy flux upper limit for each energy bin and the related systematic uncertainty.
The energy flux upper limits are shown in Fig.6. For comparison, the 1.5-year spectrum from $Fermi$ \citep{abdo11} and the solar minimum flux obtained by \citep{tang18} from $Fermi$ data are also reported along with their extrapolation to the TeV range by means of a simply power-law function. The ARGO-YBJ upper limits at sub-TeV energies are consistent with this extrapolation and close to the theoretical estimate of the maximum gamma-ray flux produced by the cosmic ray interaction with the Sun \citep{tang18,linden18}.

\subsection{ARGO-YBJ Sensitivity}

With no evident detection of gamma rays from the solar disk, we have estimated the background fluctuation and computed the expected sensitivity from off-Sun regions. Here the sensitivity refers to the average upper limit  ARGO-YBJ would  obtain in an ensemble of similar experiments with data collected from source-free regions  \citep{Feldman98}. To accomplish this task, we have applied the analysis method adopted by HAWC \citep{albert18}, therefore we have considered 64 `fake' Sun disks, 1 degree apart, outside the excluded region described in Sect.3.1, at a maximum angular distance of 21$^{\circ}$ from the true Sun position. To search for hypothetical gamma rays  the radius of the search window is chosen as in \citep{barto13}, taking into account the detector angular resolution for gamma rays shown in Table 1. In the case of purely stochastic fluctuations the distribution of the significance of the deviation between the observed counts and the expected background should be normally distributed with center at zero and unit standard deviation.

To obtain the significance, the Likelihood Ratio test has been applied
 \begin{equation}
 TS=-2lnL_0/L_1
\end{equation}

where L$_{0}$ is the likelihood for the null hypothesis (background only)  and L$_{1}$ the likelihood of the alternative hypothesis (background plus source signal). TS is expected to be asymptotically distributed as `chi-square' (dof=1) in the null hypothesis while $\sqrt{TS}$ gives the significance in units of Gaussian standard deviation \citep{mattox96}. The distribution of the significance for all energy bins is shown in Fig.5. The distribution is compatible with a normal Gaussian function, with the excess from the off-Sun regions consistent with zero.

 \begin{figure}[h!]
\centering
\includegraphics[width=3.5in,height=2.7in]{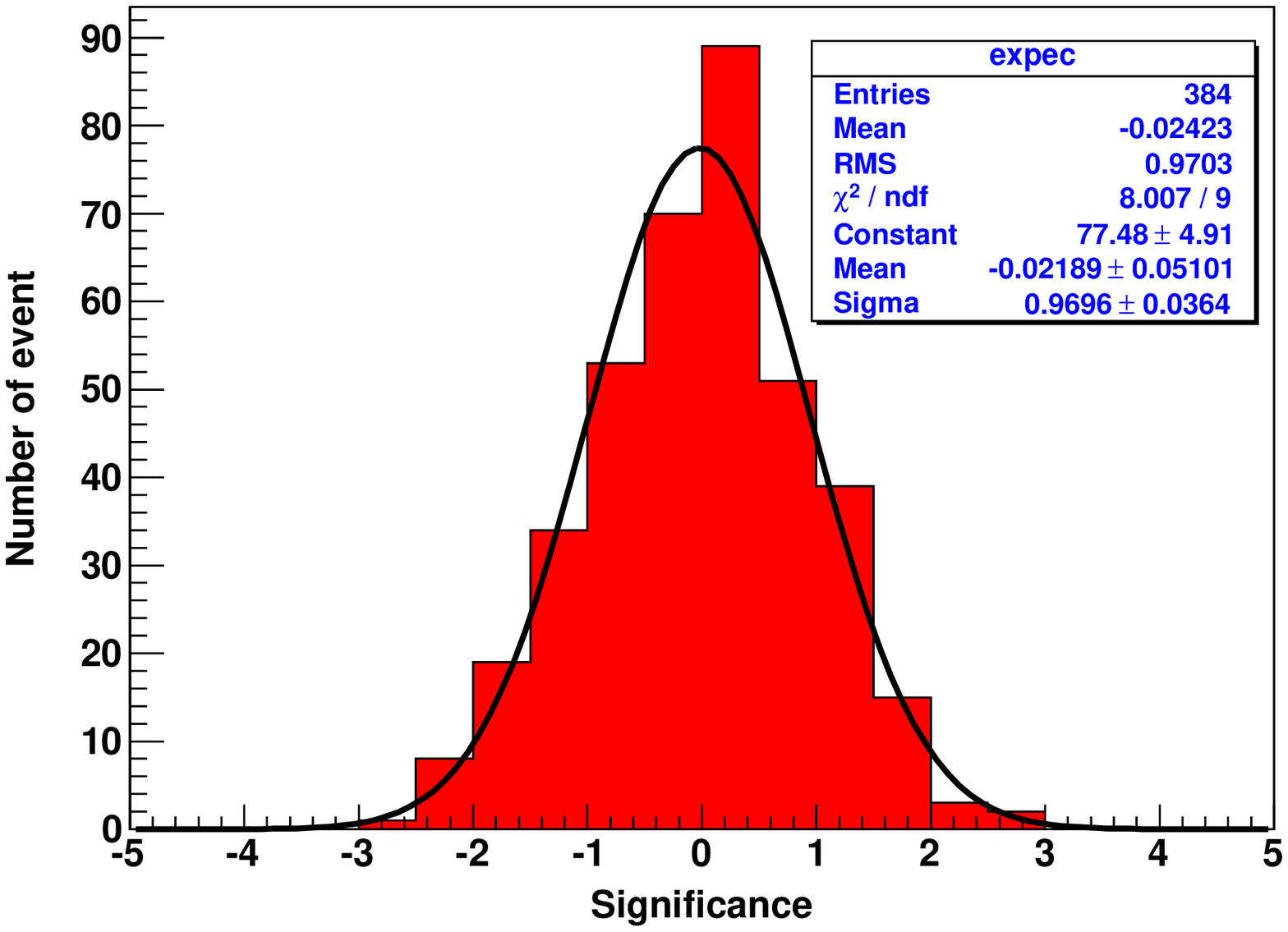}
\begin{center}
\hspace{5mm}{\textbf{Figure 5.} The distribution of the significance of the fluctuations estimated from the off-Sun regions. Data from all energy bins have been added up.}
\end{center}
\label{fig5}
\end{figure}

The 95\% upper limits on the energy flux have been calculated and their average is displayed in Fig.6 (blue squares) for each energy bin. The error bars are the standard deviation of the distributions. We observe that the sensitivity closely follows the CR upper bound spectrum. To check this result we have injected the spectrum of the CR upper bound on the background map. The significance of the excess found at the Sun's position ranges between 1.5$\sigma$ and 2.8$\sigma$, depending on the energy. The 95\% C.L. upper limits fall within the 2.5$\sigma$ range defined by the sensitivity, as shown in Table 3.

\begin{table}[ht]

\caption{The sensitivity at each energy bin compared to the 95\% upper limit on the CR upper bound spectrum injected on the background map (see text)}
\centering 
\begin{tabular}{ccccccc}
\hline
{E$_{true}$} &Sensitivity   &upper limit:CR upper bound \\
(TeV)        &(GeV cm$^{-2}$s$^{-1}$)   &(GeV cm$^{-2}$s$^{-1}$)     \\
\hline
0.28  &3.86e-8$\pm$2.09e-8  &6.58e-8   \\
0.56  &1.87e-8$\pm$8.18e-9  &3.67e-8  \\
1.01  &1.06e-8$\pm$5.04e-9  &2.31e-8  \\
2.02   &8.65e-9$\pm$2.82e-9 &1.50e-8  \\
3.76  &7.48e-9$\pm$2.80e-9  &1.08e-8  \\
6.53   &5.75e-9$\pm$2.75e-9 &8.48e-9  \\
\hline
\end{tabular}
\label{table2} 
\vspace*{0.5cm}
\end{table}

The results of this analysis prove that no anomalous fluctuation affects the background and that the ARGO-YBJ sensitivity is roughly at the level of the CR upper bound model.

The upper limits obtained from the data in low energy bins (Table 2) are consistent with the sensitivity (Table 3) which measures the ARGO-YBJ detection power. Indeed, at low energies the shadow is weak and smeared out, partially overlapping the solar disk. At high energies the strong cosmic-ray shadow coincident with the solar disk causes a substantial loss of sensitivity.

\subsection{Systematic Uncertainties}
 Systematic errors affecting the previous results are of different origin. They are mainly associated to the ROI definition, to the exposure and background estimates and,
 at a minor extent, to the assumed IC component. To ascertain the systematics related to the ROI choice, we have modified of a few degrees,
 in all directions, the limits of the ROI box. We find that the results concerning the first three energy bins may change, of 25\% for the 0.28 TeV energy bin,
 of 12\% for the 0.56 TeV energy bin, and 22\% at 1.01 TeV. The irregular shape of the Sun shadow, approximated by a two-dimensional Gaussian,
is the main source of this uncertainty. Indeed, errors less than 5\% affect the results at higher energies. Setting of the selection cuts, air shower reconstruction,
effective area estimate and background evaluation have been the subject of continuous refinements during many years of ARGO-YBJ operation. As a result,
the systematic error on the exposure is expected to be about 4\%. The systematic error on the background evaluation produces an error in the upper limits of about 13\%, 6\% and 3\%,
respectively, for the 0.28 TeV, 0.56 TeV and 1.01 TeV energy bins \citep{barto15a}. The IC component has been modeled following the calculations of \cite{zhou17}.
 Changing its normalization of 30\% produces a small effect of 2\% to the flux upper limits for the energy bins 0.28 TeV and 0.56 TeV.
 This is not surprising, being this component subdominant at the energies here concerned. The total systematic uncertainty, shown in the last column of Table 2,
 has been determined by linearly adding the individual contributions in order to get a conservative estimate.

\begin{figure*}
\centering
\includegraphics[width=7in,height=4in]{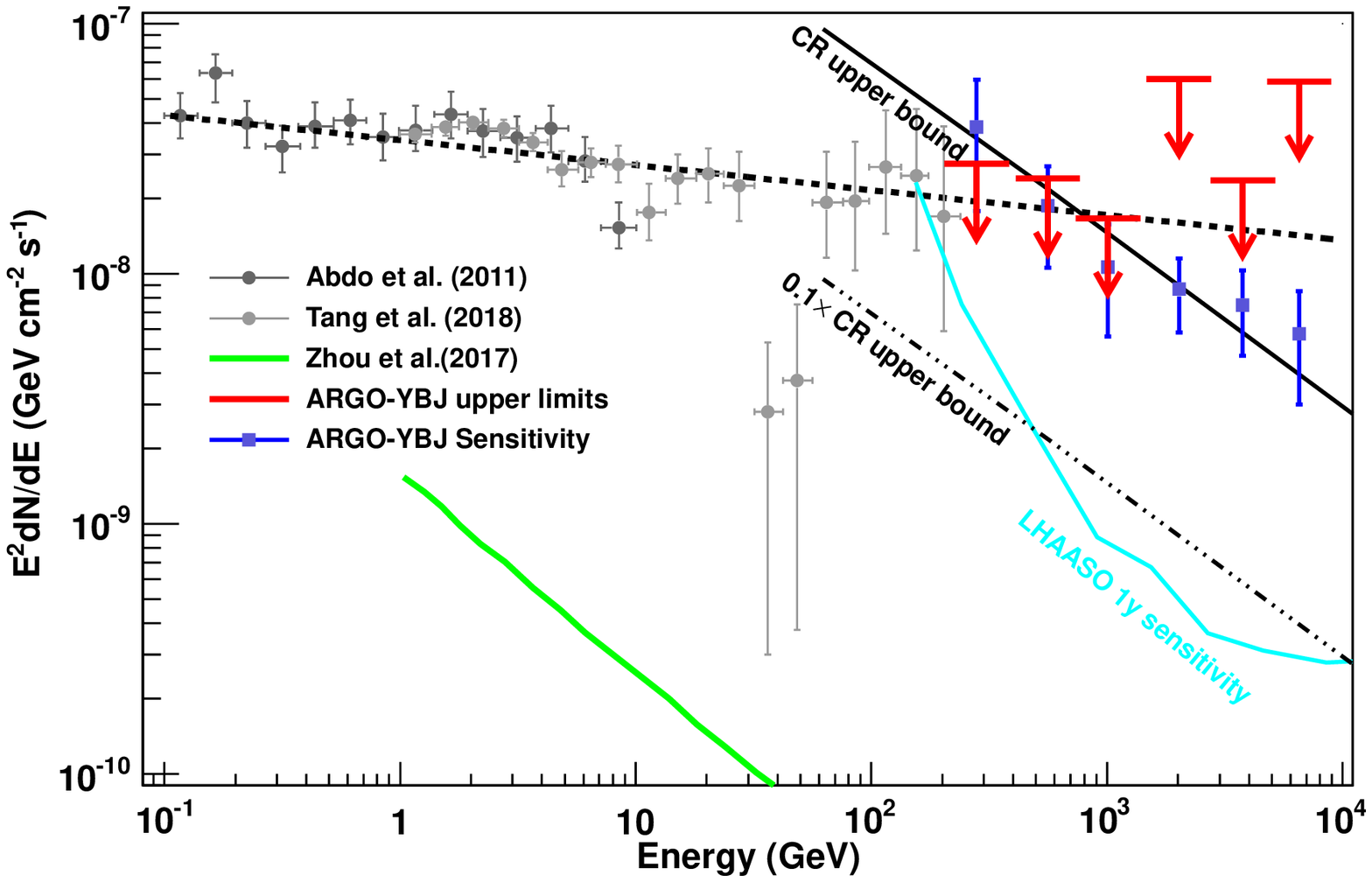}
\begin{center}
\hspace{5mm}{\textbf{Figure 6.} The solar minimum energy flux spectrum of the solar disk gamma rays measured by $Fermi$ and the ARGO-YBJ upper limits.
The data in the 0.1-10 GeV range (black dots) have been taken between 2008-8-4 and 2010-2-4 \citep{abdo11}. Data from 1 to 200 GeV (grey dots)
concern the period between 2008-8-7 and 2010-1-21 \citep{tang18}. The ARGO-YBJ upper limits come from the analysis of data collected between 2008-1-1 and 2010-12-31.
The dashed black line is the extrapolation by a simple power-law function of the $Fermi$ data to TeV energies.
The theoretical maximum gamma-ray energy flux (CR upper bound) has been calculated by \cite{linden18}.
Its 10\% scaled intensity is also shown. The theoretical lower bound of the solar disk component calculated by \cite{zhou17} without taking into account the magnetic effects is shown by a solid green line. The estimated ARGO-YBJ sensitivity at each energy bin is shown by blue squares. The 1 sigma error bars are also reported. The HAWC upper limits for 1-100 TeV gamma rays from the solar disk \citep{albert18} are near the 0.1 CR upper bound, but outside the solar minimum. For comparison, the estimated differential sensitivity of LHAASO \citep{he18} is shown.}
\end{center}

\label{fig4}
\vspace*{0.5cm}
\end{figure*}

\section{discussion}
When searching for gamma rays from the Sun using the ARGO-YBJ data, the Sun shadow may represent a disturbing offset. Indeed, the significant deficit of the Sun shadow would cover up most of the gamma-ray excess. However, as shown in Fig.3, the shadow is displaced from the Sun position due to the magnetic deflection of the cosmic rays, while the gamma-ray excess will be centered on the Sun. Since the displacement is rigidity dependent, to avoid a complete overlap of the Sun shadow with gamma rays only events with energy less than 10 TeV have been processed. On the other hand, being the displacement of the Sun shadow (Table 2) comparable with the detector angular resolution (Table 1), it results impossible to avoid a partial overlap even at low energies. For this reason the obtained flux upper limits, shown in Fig.6, are worse than that expected according to the gamma-ray sensitivity of ARGO-YBJ \citep{barto13}.

The detailed theoretical studies \citep{seckel91,zhou17} predict the solar-disk gamma-ray flux due to the cosmic ray hadrons significantly lower than the flux measured by $Fermi$-LAT up to 200 GeV \citep{abdo11,kenny16,tang18} during solar minimum. The current explanation of this discrepancy invokes the interaction of cosmic rays with the solar magnetic fields. While the cosmic ray spectrum around the Earth has been measured with sufficient accuracy, the cosmic ray flux near the Sun could be affected by these fields. An enhancement of the gamma-ray production from the solar disk is expected due to the mirroring effect on charged cosmic rays in the GeV range \citep{seckel91}. As discussed in the introduction, these effects should vanish at high energy.
Accordingly, the photon spectrum should become softer at sub-TeV/TeV energies depending on the solar cycle. Thus the extrapolation by a power-law function of the $Fermi$ results, shown in Fig.6, should provide a reasonable upper bound to the gamma-ray spectrum if this enhancement continues at TeV energies. The cosmic ray bound proposed by \cite{linden18}, shown in the same figure, represents an optimized and extreme application of the \cite{seckel91} model. This bound crosses the $Fermi$ data extrapolation at about 600 GeV suggesting that a softening of the gamma-ray spectrum should happen below 1 TeV. The ARGO-YBJ limits in this energy range, although affected by large systematic errors, are not in conflict with this hint. Thus the extension of the $Fermi$ spectrum at TeV energies would imply different physical mechanisms compared to the \cite{seckel91} model. It appears of great importance the measurement of the gamma-ray flux in the TeV range during solar minimum. The ARGO-YBJ upper limits above 1 TeV do not allow one to set any constraint on the evolution of the solar disk gamma-ray flux. The limits themselves, however, represent an experimental bound to any emission model including photon production by some unconventional process. For instance, some models which predict copious solar gamma-ray production by dark matter annihilation in the Sun via long-lived mediators \citep{arina17,leane17} are already excluded by the ARGO-YBJ data. Indeed, if the mediators have a long lifetime leading to a decay length greater than the solar radius, an enhanced flux of photons is produced. A detailed study of the portion of the model parameter space compatible with the ARGO-YBJ bound is beyond the scope of this paper. This study has been carried out in the TeV range by the HAWC collaboration \citep{albert18a} providing the strongest existing constraints on the spin-dependent cross-section of TeV dark matter with protons. We just point out that an energy flux exceeding 10$^{-7}$ GeV cm$^-$$^2$ s$^-$$^1$ at sub-TeV energies is ruled out by the ARGO-YBJ limits.

\section{Summary}

The gamma-ray emission from interactions of charged cosmic rays in the solar atmosphere has been firmly detected by the $Fermi$-LAT detector up to 200 GeV during solar minimum at the end of Cycle 23. The measured flux is significantly higher and the spectrum shape flatter than predicted. The origin of these discrepancies is currently attributed to solar magnetic effects which are expected to end at higher energies. Motivated by these findings, we have searched for solar gamma-ray emission at higher energies using 3 years data recorded by the ARGO-YBJ experiment during the solar minimum from January 2008 to December 2010. Upper limits have been set to the solar disk flux in the energy range 0.3-7 TeV . The corresponding energy fluxes are 1.6-6.0$\times$10$^-$$^8$ GeV cm$^-$$^2$s$^-$$^1$, consistent with the naive extrapolation of the $Fermi$ data. This result represents the first observational study of the solar disk gamma-ray emission in the TeV range during solar minimum. The ARGO-YBJ data do not provide any clear answer to the question concerning the extension of the solar magnetic effects at high energies, though suggesting a softening of the gamma-ray spectrum before 1 TeV. Dark matter models predicting energy fluxes exceeding 10$^-$$^7$ GeV cm$^-$$^2$s$^-$$^1$ at sub-TeV energies are strongly constrained. The limited sensitivity of ARGO-YBJ is mainly due to the presence of the Sun shadow whose displacement is comparable to the detector angular resolution. This problem should be at least partially solved by EAS arrays as HAWC \citep{abey13}, currently running, and LHAASO \citep{zhen14}, in construction, owing to their capability to reject a large fraction of the hadron-induced air showers. The sensitivity predicted for LHAASO \citep{he18} is shown in Fig.6. HAWC is
expected to provide comparable sensitivity \citep{albert18}. Whether this sensitivity would be properly achieved, the detection of solar gamma rays could provide new insights about the modulation of cosmic rays in the solar magnetic field and their interaction with the solar atmosphere, and probe models of dark matter annihilation in the Sun. It is worth to note that if the cosmic ray background cannot be completely rejected, a residual Sun shadow of reduced depth will still remain. In this case, the approach used in this work could usefully be adopted.

\acknowledgments
We are grateful to the anonymous referee who helped us to improve this
paper considerably. This work is supported in China by NSFC (No. 11575203, No. 41504080),
the Chinese Ministry of Science and Technology, the
Chinese Academy of Sciences, the Key Laboratory of Particle
Astrophysics, CAS, and in Italy by the Istituto Nazionale di Fisica
Nucleare (INFN).
We also acknowledge the essential supports of W. Y. Chen, G. Yang,
X. F. Yuan, C. Y. Zhao, R. Assiro, B. Biondo, S. Bricola, F. Budano,
A. Corvaglia, B. D'Aquino, R. Esposito, A. Innocente, A. Mangano,
E. Pastori, C. Pinto, E. Reali, F. Taurino and A. Zerbini, in the
installation, debugging and maintenance of the detector of ARGO-YBJ.


\clearpage

\end{document}